\documentstyle[11pt,aaspp4]{article}
\slugcomment{ApJ, submitted}
\lefthead{Andreon \& Cuillandre}
\righthead{}
\begin{document}
\input psfig.tex 
 
\title{Dissecting the luminosity function of the Coma cluster of galaxies
using CFHT\footnote{Based on observations obtained at the 
Canada--France--Hawaii Telescope
and in part at the {\it Hubble Space Telescope}} 
wide field images}

\author{S. Andreon} 
\affil{Osservatorio Astronomico di Capodimonte, via Moiariello 16, 80131
Napoli, Italy \\ e-mail: andreon@na.astro.it}

\and
\author{J.-C. Cuillandre}
\affil{Canada--France--Hawaii Telescope, 65-1238 Mamalahoa Hwy., Kamuela, 96743, Hawaii\\ 
e-mail: jcc@cfht.hawaii.edu}
 
\begin{abstract}

We determined the relative spatial density of the Coma cluster galaxies
selected by luminosity, and the contribution of the galaxies of each
central brightness to the luminosity function (i.e. the luminosity
function bi--variate in central brightness). The Coma cluster and 
control fields were imaged using the CFH12K ($42\times28$ arcmin)
and UH8K ($28\times28$ arcmin) wide--field cameras at the 
Canada--France--Hawaii Telescope. Selected {\it Hubble Space Telescope} 
images were used for testing.

Quantities were derived from measurements in at least two colors, which 
have the following features: (1) Galaxies as faint as three times the 
luminosity of the brightest globular clusters are in the completeness region 
of our data. (2) We have a complete census (in the explored region) of low 
surface brightness galaxies with central surface brightness galaxies almost 
as low as the faintest so far cataloged ones. (3) The explored area is among 
the largest ever sampled with CCDs at comparable depth for any cluster of
galaxies. (4) The error budget includes all sources of errors known to
date. Using {\it Hubble Space Telescope} images we also discovered
that blends of globular clusters, not resolved in individual components
due to seeing, look like dwarf galaxies when observed from the ground and
are numerous and bright. When mistaken as extended sources, they
increase the steepness of luminosity function at faint magnitudes. The
derived Coma luminosity function is relatively steep ($\alpha=-1.4$) over
the 11 magnitudes sampled, but the slope and shape depend on color. A
large population of faint low surface brightness galaxies was discovered,
representing the largest contributor (in number) to the luminosity
function at faint magnitudes.  We found a clear progression for a
faintening of the luminosity function from high surface brightness
galaxies ($\mu\sim20$ mag arcsec$^{-2}$) to galaxies of very faint
central brightnesses ($\mu\sim24.5$ mag arcsec$^{-2}$), and some evidence
for a steepening. Compact galaxies, usually classified as stars and
therefore not included in the LF, are found to be a minor population in
Coma.

\end{abstract}
 
 
\section{Introduction}
 
The luminosity function (LF hereafter), i.e. the number density of galaxies
having a given luminosity, is critical to many observational and theoretical
problems (see e.g. Binggeli, Sandage \& Tammann 1988). From an observational
point of view, the LF is the natural ``weight" of all those quantities which
need to be weighted against the relative number of objects in each
luminosity bin. Furthermore, due to the roles played by flux and surface
brightness in the inclusion of objects in any observed sample (faint objects
or low surface brightness galaxies are often excluded or
under--represented), the knowledge of the LF and the LF bi--variate in
surface brightness is fundamental to compute the selection function and is
needed to derive the actual galaxy properties from the measured quantities
(see, for example, the discussion on the field LF steepness by Sprayberry et
al. 1997). 

The optical LF of galaxies in clusters has been extensively studied (e.g.,
to cite just a few papers dealing with large number of clusters, Gaidos
1997; Valotto et al. 1997; Lumsden et al. 1997; Garilli, Maccagni \&
Andreon, 1999). However, faint dwarfs and low surface brightness galaxies
are outside the reach of most of the previous investigations. Furthermore,
the existence of compact galaxies is usually ignored, because, in practice,
they are misclassified as stars and then removed from the sample
(see, as an exception Drinkwater et al. 1999). Therefore, an extension of
the LF to fainter magnitudes and lower surface brightnesses, and without any
assumption on the compact galaxy contribution, and possibly bi--variate in
surface brightness would be profitable.

Most importantly, the {\it global} LF hides the true problem
(Sandage 1994): the LF is the sum of the LFs of the specific types, or of any
other physically based galaxy classes. In fact, the LF is dependent on the 
environment as shown by Binggeli, Sandage \& Tammann (1988). Maybe the LFs of
the morphological types are universal (Binggeli, Sandage \& Tammann 1988,
Jerjen \& Tammann 1997, Andreon 1998), but faint galaxies whose morphological
type do not fit well in the Hubble (1936) morphological scheme (which has been
built for classifying giant galaxies, not dwarfs) raise some concern on the
extension of the {\it type--dependent} LF at faint magnitudes. Galaxies can be
also classified on the basis of their central brightness which also determines
where they fall in the Fundamental Plane
(e.g. Bender, Burstein \& Faber 1997), showing that this classification 
reflects some physical difference between the classes. Therefore, ``it would be
of great importance to know what the luminosity function looks like when
divided into classes of surface brightness'' (Kron 1994). In contrast to Hubble
types, classes of central brightness are continuous (as often nature is),
quicker to determine, and can be computed with observations of lower quality
than those required to determine morphological types. However, brightness
classes merge in the same class giant galaxies of the different morphological
types, which are known to have different properties (see, e.g. Andreon 1996 for
Coma galaxies).

\medskip 

In the present paper we present in three colors the LF and in two colors
the LF bi--variate in surface
brightness of a sample of galaxies in the Coma cluster of about 1000
members. We measure them down to the magnitude of three
bright globular clusters, and to the brightness of the faintest cataloged low
surface brightness
galaxies. Our studied area is among the largest cluster area ever
observed with CCDs. We use the standard method for computing the LF, namely,
the method of differential counts (Zwicky 1957). The method is quite simple:
the LF  of the cluster galaxies is the difference between galaxy counts in the
cluster direction and those counted in a control field direction devoid of
(cataloged) clusters.  This method has a some advantages: 1) It does not
require an extensive redshift survey. 2) The redshift dependence of the K
correction is not needed.  3) The number of galaxies in each magnitude bin is
proportional to the natural frequency with which galaxies are found in the
Universe, at least in clusters.  4) The difficult problem of calculating
the visibility function for a mixed diameter+flux limited survey (as all {\it
field} survey actually are) is completely skipped, because the cluster sample
is naturally volume--limited (details are presented in Section 3). The method
has the main shortcoming that it applies only to galaxy over--densities, and
that galaxies in clusters could not be representative of galaxies in general.
In that case, the study of cluster galaxies could reveal a correlation
between the cluster environment and galaxy properties.

For the Coma cluster, we adopt distance modulus of 35.1 mag (i.e. $H_0=68$ km
s$^{-1}$ Mpc), according to the direct measure by Baum et al. (1997). The
slope of the LF is, as for the Schechter (1976) function, defined by

$$\alpha=-\frac{1}{0.4} \frac{ \partial \log LF }{ \partial m} -1$$

in such as way that a flat (in mag) LF has $\alpha=-1$.

\section{The data}

$B$, $V$ and $R$ Coma cluster observations were taken on January 12th, 1999,
during the CFH12K (Cuillandre et al. 2000) first light at the
Canada--France--Hawaii telescope prime focus in photometric conditions. Table 1
summarizes a few relevant characteristics of the observations. CFH12K is a 
12,288 $\times$ 8,192 (12K$\times$8K) pixel CCD mosaic camera, 
with a 42 $\times$ 28 arcmin$^2$ field of view
and a pixel size of 0.206 arcsec. The four dithered images per filter were
pre-reduced  (overscan, bias, dark and flat--field) and then optimally
stacked. The CFHT
CCD mosaic data reduction package FLIPS (Cuillandre 2001) was used.  Figure 1
shows the studied field. For the present scientific analysis of these very
early observations, only the best part of the image is kept (3 low grade CCDs
were replaced a few months later), consisting of $\sim10.8$ CCDs in $V$ and $R$
(1.2 CCDs are of engineering quality) and 8 CCDs in $B$ (1.8 more CCDs are
partially vignetted by the only $B$ filter available during these early
observations). After discarding areas noisier than average (gaps between CCDs,
borders, regions near bright stars and large galaxies, etc.), the usable area
for the Coma cluster is 0.29 square degrees in $V$ and $R$ and 0.20 degrees sq.
in $B$. Images were  calibrated in the Bessel--Cousin--Landolt system through
the observation of photometric standard stars listed in Landolt (1992). The
scatter of the zero--point measured for the sub-sample of 7 to 12 individual
stars with large $m$ in the Landolt (1992) catalog (i.e. observed during
several nights  by him) and in the field of view of the images, is
$\sim0.02-0.03$ mag, in the three filters. We do not find any trend for a
zero--point dependency on magnitude, color, CCD considered, and apparent
location in the field of view. Photometric calibration has been cross checked
by using aperture magnitudes of a few galaxies in our field of view listed in
de Vaucouleurs \& Longo (1988). This external check rules out zero--point
errors larger than $\sim 0.1$ mag.

The $B$ band control field is the area around the galaxy NGC\,3486 (which
occupies less than 10 \% of the camera field of view, a $10 \time 4$
arcmin$^2$ area). This field shares the photometric calibration of Coma 
and we have checked the photometric
zero point, at a 0.1 mag error level, by comparing our aperture photometry
of NGC\,3486 with that listed in de Vaucouleurs \& Longo (1988). $V$ and
$R$ control field images (SA 57) have been taken from the archive of one
of us (J.-C. C.). They were taken in 1998 at the same telescope, through
identical filters, but with the UH8K camera equipped with frontside
illuminated CCDs. UH8K is an 8K $\times$ 8K mosaic camera with a 28
$\times$ 28 arcmin$^2$ field of view and a pixel size of 0.206 arcsec.
This SA\,57 field is centered on a region devoid of (cataloged) clusters,
and includes a photometric sequence (Majewski et al. 1994), which allows
an accurate and straightforward photometric calibration. No significant
color term has been detected (as none is present in the CFH12K images).
One of the CCDs of the UH8K presented a severe charge transfer problem,
and for simplicity it has entirely discarded from further analysis. These
images cover a large area of $\sim650$ arcmin$^2$ and they are angularly
distant enough from the Coma cluster (a bit more than 2 degree,
corresponding to 3.4 Mpc, or a 1.5 Abell radii at the Coma cluster
distance) not to be strongly contaminated by its galaxies, but near enough
to sample the overdensity associated to the Coma supercluster. 
However, our $B$ band control field samples a background several degrees away from the
Coma cluster direction. Field images were processed following the same
procedure applied to the Coma cluster data. 

Objects are detected using SExtractor (Bertin \& Arnouts 1996), using
standard  settings (a minimal area of four pixels and a threshold of $\sim
1.5 \sigma$ of the sky).

\section{Method of differential counts}

The cluster LF (or, equivalently, the relative space density distribution of
galaxies of each luminosity) is computed as the difference between galaxy
counts in the Coma and in the control field directions (for an introduction on
the method, see, e.g., Oemler 1974). The LF bi--variate in central brightness
is computed in a similar way by subtracting off the contribution due to
foreground and background measured in the control field from counts in 
the Coma direction. 

The method is robust, provided that all sources of errors are taken
in to account. Several of them have already summarized in Bernstein et 
al. (1995), Trentham (1997) and Driver et al. (1998) and not repeated here.
We remind that:

-- Extensive simulations show that undetected galaxies cannot
be confidently recovered, even statistically, so that completeness
corrections are unreliable (Trentham 1997). Therefore, it is preferable,
as we did, to cut the sample at the magnitude of the brightest galaxy of
the faintest detected surface brightness.

-- Gravitational lensing distorts background counts in the cluster line of
sight (Bernstein et al. 1995; Trentham 1997) but it is negligible in very
nearby clusters, such as Coma.

This is a quickly growing field, so one should be aware that terms
that presently are currently included in the error budget, such as
non--Poissonian fluctuations, were not included just few years ago (but
there are exceptions, such as Oemler 1976).

A few sources of concern that should be considered in computing the LF and bi--variate 
LF of galaxies are discussed in more detail in the next sections.

\subsection{Nearby background}

By using a control field that crosses the Coma supercluster (in $V$ and
$R$), we are able to measure the Coma cluster LF without the contamination
of the large scale structure in which it is embedded, unlike almost all
previous CCD determinations of the cluster LF that used control fields in
areas too distant from the studied cluster (e.g. Bernstein et al. 1995;
Trentham 1998a; Lobo et al. 1997; Biviano et al. 1995). It is easy to show
(Paolillo et al. 2001) that a control field too close to the cluster,
and therefore contaminated by cluster galaxies, does not alter the shape
of the LF, but just changes the LF normalization (and makes errorbars
larger) if the LF does not depend too much on environment.

\subsection{Photometric errors}
 
The photometric quality of the night, our checks with aperture photometry
of catalogues galaxies and the presence of photometric standard in the
field of view, all exclude photometric errors as significant source of
errors in the determination of the LF or the bi-variate LF.

\subsection{Background errors}

Fluctuations of galaxy counts are surely no longer simply Poissonian in nature
(i.e. due only to small number statistics), because of a non--zero correlation
function, or, in simple words, because of the existence of clusters, groups and
voids. The fluctuation amplitude can be directly measured, as in Bernstein et
al. (1995), or estimated by using the Huang et al. (1997) formalism, which uses
the galaxy angular  correlation function in order to estimate the galaxy count
variance  averaged over the field of view at a given magnitude and   passband.
We use this last method, owing just one control field.  Background fluctuations
is, in most of the luminosity bins, the largest term in the error budget.
Huang et al. (1997) provide the amplitude of the background variance in a
given magnitude bin and in a given area, once the characteristic ($M^*$)
luminosity  of the field population is given, by adopting a galaxy--galaxy
spatial correlation function. As characteristic luminosity, we adopt $B=-20.5 \
, V=-21 \ , R=-21.7$ mag (Zucca et al. 1997;  Garilli, Maccagni \& Andreon
1999; Paolillo et al. 2001; Blanton et al. 2001). Adopting characteristic
luminosities that differ by up to one mag does not change appreciably the
errors.

Errorbars for the bi--variate LF further assume (because of the lack of
appropriate measures) that the correlation scale of the galaxy angular
correlation function is the same for all galaxies, independently on their
central brightness. Thus, errorbars are approximate, but we verified that a
difference in the clustering scale of a factor two produces negligible changes
to our results. 

Due to the fact that the LF is the difference between ``cluster+background''
and ``background'', the error on the LF has two terms related to the background.
In almost all literature LFs, only one term related to the 
background is taken into account, under the implicitly assumption that the
``true'' background counts are perfectly known. 

\subsection{Adopted magnitudes and Low Surface Brightness Galaxies}

Visual inspection of our images shows that several faint objects in the Coma
direction are larger, when measured at $\mu\sim25$ mag arcsec$^{-2}$, than
those in our control field, where most of the faint objects are small. The
adopted detection thresholds ($\mu=25.0,25.5,24.5$ mag arcsec$^{-2}$ in $B,
V$ and $R$, respectively) are fainter than the typical central brightness of
low surface brightness galaxies (LSBGs, hereafter), which range from 22 to 24
$B$ mag arcsec$^{-2}$ (McGaugh, Schombert \& Bothun 1995; Bothun, Impey \&
McGaugh 1997). Quite recently (O'Neil et al. 1999) LSBGs of central
brightness as faint as $\mu_B=24.5$ have been counted.


Therefore, our detection threshold is as low as, or just slightly brighter
than, the lowest central brightness sampled so far, with the notable
exception of Ulmer et al. (1996) LSBGs. 

The measured luminosity of LGBGs
is strongly dependent on the integration radius because of their shallow
surface brightness profiles. We adopt isophotal magnitudes, recognizing
that these magnitudes include a fraction of the object luminosity
depending on the object central brightness and on the radial surface
brightness profile. Our magnitudes are not, therefore, {\it total}
magnitudes. In Sect 5.1 we discuss the impact of this choice on the LF. 

Galaxy counts are strongly dependent on the type of magnitude (aperture,
isophotal, asymptotic, etc) used for measuring the flux, and in the
cluster direction this effect is exacerbated by nearby (and therefore
large) galaxies.  Our field counts agree with those in literature (Driver
et al. 1994; Trentham et al. 1998) once we select the same type of
magnitude adopted in the comparison work. We found that galaxy counts are
significantly lower when the adopted isophotal magnitudes are used. We
notice that galaxies with normal colors are easier to detect in $V$ and
$R$ than in $B$, because of the much brighter detection threshold in the
latter filter. 

\subsection{Completeness}

Since undetected LSBGs can not be recovered, we need to cut the sample at
the magnitude of the brightest LSBGs of the faintest detectable
central surface
brightness. A detailed explanation of this method is described in Garilli
et al. (1999). By definition, the sample will be complete down to the
cutting magnitude. For our sample, the cutting magnitudes are:  $R=23.25,
V=23.75$ and $B=22.5$ mag. At these magnitudes the measured signal
to noise ratio is about 20. 

\subsubsection{LSBGs} 

Due to the low surface brightness threshold, LSBGs are included in our
catalogue. Galaxies with extremely low central surface brightness ($\mu_0 >
\sim 25$ mag arcec$^{-2}$) are correctly excluded in our LFs
because their magnitude at the chosen isophote is exactly zero.

\subsubsection{Eddington bias}

Catalogues suffer a usual incompleteness: due to the noise, galaxies can be
undetected even if their central brightness is slightly brighter than the
threshold, and can be detected even if their brightness is below the
threshold. Furthermore, the noise and the increasing galaxy counts at
faint magnitudes include in catalogues a larger number of galaxies than
they exclude (this effect is called Eddington bias). By keeping only high
quality data, as we do by cutting the samples at the completeness
magnitude, incompleteness and Eddington bias are a minor concern. For
example in our fainter bin, the observed minimal signal to noise
ratio (S/N) is $\sim20$ in $R$, while at the faintest magnitude and at the
faintest surface brightness, the observed S/N of the central
brightness is $\sim10$ in $R$. 

\subsection{Image properties matching}

The control field images are deeper than Coma images, and taken under
better seeing conditions, with the exception of the $B$ images that were
taken during similar seeing conditions (see Table 1). In order to compute
the LF and the bi--variate LF, it is necessary to match the properties of
the control and program images. First of all, we match the seeing profile,
convolving control field images with an appropriate kernel. The match of
point spread function is checked by verifying that stars lay on the same
magnitude $vs$ central brightness locus, in both the Coma and the control
field images. Then, the noise in the images is matched by adding
Poissonian noise. We checked that the noise matching is not crucial, i.e.
the results do not change by more than the errorbars. This holds because
we take the general approach of completely discarding all data which are
affected by noise. By cutting our sample to a minimal
signal to noise of 20, noise is not a concern.

\subsubsection{Star/galaxy classification and compact galaxies} 

Careful numerical simulations performed by us show that existing
elliptical galaxies as compact as NGC\,4486B or M\,32, could not be
recognized as galaxies in our images {\it independently of their
luminosity} if they were in the Coma cluster\footnote{This concern has
been raised by Dave Burstein, that we warmly thank.}, and they look like
stars on our images. 

As previously stated, the LF is given by the difference of {\it galaxy}
counts. What is actually usually taken in literature is the difference of
counts of {\it extended} objects. The two calculations give the same
result when galaxies and ``extended objects'' classes perfectly overlap,
however this hypothesis is not satisfied even in a cluster as near as the
Coma one. 



Excluding {\it ab initio} compact galaxies from the class of galaxies,
they could not be counted in the LF.

How to solve this problem? In two ways, depending on the object
luminosity:

a) Bright objects. Our control field is close enough in the sky to the
Coma cluster to assume that star counts are equal, within the statistical
fluctuations, in the two pointings (which are both at the Galactic Pole
and whose nearest corners are less than 1 degree apart).  We verified by
means of Besan\c con models
(http://www.obs-besancon.fr/www/modele/modele.html) that the variation of
star counts due to the small differences in Galactic latitude and
longitude between Coma and the control field is negligible (far less than
1\%).  We can check the existence of bright compact galaxies
(misclassified as stars) by simply comparing the number of the star--like
objects in the Coma and control field directions. In the control field
there are 236 objects brighter than $V=20.5$ mag classified as stars. The
expected number of stars in the Coma pointing (which covers a larger area)
is thus 384. We found 382 stars, two less than the expected number, and
therefore no excess of compact objects in the Coma direction is found.
The $1\sigma$ upper limit to the number of compact ellipticals in the
studied portion of Coma is 25. Even if these 25 galaxies were present
(while we found $-2$ galaxies), they are a minor
population (a 9 \% of the net number of Coma galaxies brighter than
$V=20.5$ mag) and they change the measured Coma cluster FL by less than 
errorbars.

Due to the verified paucity of compact galaxies in Coma, bright
stars (brighter than $V=20.5$ mag) are individually removed from galaxy
counts. Unlike previous works, we have verified that compact galaxies are
a minority population before discarding them.

b) Faint objects. At faint magnitudes, even not so compact galaxies can be
misclassified as stars due to noise in $V$ and $R$. In fact, we found that
several objects from the control field are misclassified at $V>21$ mag, when
the images are degraded to match the seeing and noise of Coma ones.
Furthermore, star counts differ in the Coma and field directions at faint
magnitudes (but not at bright magnitudes), whereas they should
be equal according to the model. Therefore, stars are not individually
identified and removed, but statistically subtracted and star--like
objects in the Coma direction due to compact galaxies are not thrown away
during the star/galaxy classification. As a consequence, the problem of the
star/galaxy misclassification (both due to object faintness and to the
intrinsically object compactness) is overcome. This way, the problem
represented by compact objects is solved, but at the price of larger
errorbars because of the statistical subtraction. We stress out that {\it
measured} stars counts are used, not the expected ones.

\subsection{Globular clusters and their blends}

Even a casual inspection of the region around IC\,4051, an early type galaxy in
the studied field shown in the bottom panel of Figure 1, shows a huge
population of extended sources clustered around this galaxy. Other extended
sources are present near NGC\,4481, another bright Coma elliptical in our field
of view.  These objects are extended and as bright as $R=21$ mag. Since
globular clusters (GCs hereafter) of IC\,4051 have a turnoff magnitude of
$V\sim25$ (Baum et al. 1997), and  are unresolved at the Coma distance (i.e.
they are point sources), these huge population can not be formed by individual
GCs.  In order to understand how many extended source there are at each
magnitude, we compute their luminosity function. We first subtract a model of
the galaxy, obtained by fitting its isophotes. Then, we compute the counts in
an annulus centered on IC\,4051 of 6 and 31 arcsec of inner and outer radii
respectively, and in a control region of the same area at 160 arcsec East of
IC\,4151.  In the annulus on IC\,4051 we found an excess of $2.3 \times 10^5$
too many extended objects per mag per square degree at $R\sim24$, with respect
to the control field. The number of extended objects in the annulus is four
times larger than in the control field, and the excess is statistically
significant, even including non--Poissonian fluctuations.  The luminosity
function of these extended sources have a slope, in a 3 mag magnitude range
fainter than $R=21$ mag, compatible with the slope of the GCs specific
frequency ($0.4$).   The brightest of these extended sources has $R=21$ mag,
i.e. they are  $\sim6$ mag brighter that the GC turnoff (directly measured by
Baum et al. (1997) for this galaxy), and 3.4 mag brighter than the tip of the
GC population (which in turn is ill defined, because the number of bright GCs
decrease exponentially at bright magnitude without any clear break).

Since IC\,4051 has been observed by the {\it Hubble Space Telescope} (Baum et
al. 1997) we can use the superior angular resolution of the {\it HST} for
better understanding these sources. {\it HST} archive images of IC\,4051 have
been retrieved, the galaxy has been modeled and subtracted off, as for the
ground images. Figure 2 shows the residual image of IC\,4051, as seen in our
ground image (left panel) and from the space (right panel). In the left panel,
the actual galaxies revealed by {\it HST} are marked by circles. Notice that
only one faint object is circled. All the other objects are blends of a few
point sources (typically three to five), unblended at the {\it HST} resolution.
Most of them are brighter than our (and other deep probing of the LF)
completeness limit, and therefore would be counted as galaxies in the LF. The
two  brightest blends in the {\it HST} field of view have $R=21.3$ and $R=20.8$
mag. The large majority of {\it HST} point sources are GCs (Baum et al. 1997),
and therefore the large majority of our extended sources are blends of GCs.
However, a few extended sources could be blends of any type of point sources,
such as foreground stars, GCs and groups of GCs if they exist, because even
{\it HST} cannot individually distinguish GCs, at the Coma distance, from
foreground stars. In particular, the two brightest sources marked with a
diamond in the {\it HST} image are largely dominated by a bright single point
source, quite bright to be a single GC, whose identification as GC or
foreground star is possible only on statistical basis. 

Simple statistical
arguments on the luminosity function of GCs suggest that the
very brightest of our blends are blends of GCs and any other source
unresolved at the {\it HST} resolution (including groups of GCs if they
exist), while the other ones are instead, in
large majority, blends of GCs alone.

Inspection of the {\it HST} image of another large galaxy in our field,
NGC\,4481 (Baum et al. 1995), confirms our findings for such blends.

To summarize, our large population of extended sources are, in large majority,
blends of GCs. While GCs are point sources, their blends are a source of
concern because they have the unfortunate property of being classified as
single extended sources in typical seeing conditions, and thus are included in
the galaxy counts. Being blends of a few/several GCs, these sources are
brighter, on average, than GCs.  Therefore, GC blends do not only affect GC
typical magnitudes ($V\sim27$ mag), but also bias much brighter counts (as
bright as $R=21.5$ mag) and are thus pernicious because they are extended
sources. Their density is high near giant ellipticals, four times higher than
galaxy counts in the considered region of IC\,4051.

Previous works studying the deepest part of the galaxy LF may be affected by GC
blends at faint magnitudes. For example, the Bernstein et al.'s (1995)
determination of the Coma LF at very faint magnitudes, measured in the
NGC\,4874 outer halo, optimistically assumes that the GCs contamination starts
at $R=23.5$ mag (while it starts at 2 magnitudes brighter) and rule out a GC
contamination at brighter magnitudes because their objects are marginally
resolved, while we found that also GC blends share this property. De Propris et
al. (1995) found a steep LF over their very small studied field (with a slope
that nicely corresponds to those of our GG blends), and they correctly warn the
reader on the possible contamination of their galaxy counts by an unusual
population of GCs. Actually, we believe that their counts are contaminated by
GC blends more than an unusual population of GCs, because of the similarity of
the properties of their possible unusual population of GCs to our GC blends and
because Trentham (1998) does not find such a steep slope when observing one the
De Propris et al. (1995) clusters over a larger field of view (where the
contribution of GC blends is washed out). 

Thus, the points of published LFs at $M>\sim-14$ mag should be regarded with
caution as long as the area surveyed is comparable (or smaller) than that
occupied by bright galaxies. Because of this potential source of error, we
generously mask out areas to discard  a few bright galaxies with a large GC
population and a halo. Residual unflagged contamination is diluted by the very
large field of view of our images. Flagging areas occupied by large galaxies
also solves in the simplest way the problem of crowding, because the unflagged
area is mostly un--crowded.

\section{The Coma cluster LF, the bi-variate LF}

With respect to previous LF determinations, our work presents new features: 

-- The control field, although only a single one, is at an ideal angular
distance from the cluster pointing: far enough from the Coma cluster not
to be strongly contaminated by its galaxies, but near enough for correctly
sampling the density enhancement of the Great Wall (in which the Coma
cluster is embedded). Even if the control field were contaminated by Coma
cluster galaxies, the shape of the LF would not be altered by this
contamination. Background fluctuations are included in the error
budged.

-- Compact galaxies are not lost in the star/galaxy classification, and no
assumption about their existence, or contribution to the LF, is made. 

-- We do not assume that galaxy counts in the control field are the ``true''
average errorless background, and in measuring errorbars, we count twice
background errors.

-- Blends of GCs are not counted in galaxy counts.

\subsection{Luminosity Function}

Figure 3 shows (filled points) the Coma cluster LF down to
$R=23.25, V=23.75$ and $B=22.5$ mag. Notice the large number of
galaxies per magnitude bin in our $R$ and $V$ LFs and the absolute faintness
of studied galaxies ($M_R\sim-11.75, M_V\sim-11.25$, and $M_B\sim -13$ mag),
whose luminosity exceeds the tip of the GC LF ($M_V\sim-10$ mag) by less than
a factor 3 in flux (in the deepest bands). The LF extends over an 11
magnitude range and it is one of the deepest ever derived from
CCD photometry. 

The LFs in the three filters present both similarities and differences.
The LFs seem truncated at the bright end ($R=12$, $V=13.5$, $B=15$ mag).
This abrupt truncation is due to the fact that all galaxies  brighter
than the first plotted point are
removed from the sample because their potential large population of GCs
(and their blends). 

At intermediate luminosities ($B<18, V<16$ and $R<16$ mag) the LFs are fairly
flat. 

At fainter magnitudes, the LFs are steep in
$R$ and $V$, and with a much shallower slope in $B$. Of course, the exact
slope depends on the considered magnitude and filter and can be precisely
computed by the reader at his favorite magnitude by taking the best fit
functions whose parameters are listed in Table 2 
\footnote{There are
shortcomings in inferring a slope from a parametric fit to the data, see,
e.g. Merritt 1994, when the terms ``slope" indicates the local derivative
of the underlining function in a given point. 
However, the latter is not the usual meaning done to the term ``slope"
which instead means the typical change of the LF over a finite
magnitude range, as the Schechter and our power law functions provide.}
or by using the tabulated LF included as electronic table.
The typical slopes range from $-1.25$ in $B$ to $-1.4$ in $R$ and $V$.  
In the three
filters we do not see any clear turn off of the LF, meaning that galaxies
can be as faint as 3 very bright GCs and such galaxies are the most numerous in
the studied Coma region. In the $V$ band there is a
hint of a flattening of the LF at faint magnitudes, but the statistical evidence
for it, or for a turn off of the LF, is small due to the large errors. 

The quality of our LFs decreases going toward blue filters for two
reasons: first of all, the surveyed area in $B$ is
30 \% smaller than in $V$ or $R$. Second, bluer filters select preferentially
blue galaxies, abundant in the field and rare in clusters and therefore the
contrast between members and interlopers is low. Because of these reasons, 
the bi--variate LF in the $B$ band is not presented.

In the $R$ band, the LF shape is not well described by Schechter (1976)
law, because their best fit has $\chi^2\sim37$ for 18 degrees of freedom. A
function with more free parameters better describes the data. 
The best fit with a 3$^{rd}$
order power--law (i.e. with one more free parameter) is overplotted in
Figure 3 (smooth curves). The best fit parameters are listed in Table 2. The
reduced $\chi_{\nu}^2$ is $\sim 1$, suggesting a good fit.

The LF does not continue to steepen any more at longer wave bands, because
in the $H$  ($\lambda \sim 1.6 \mu m$) band, the LF of the same
portion of the cluster has slope $\alpha=-1.3$ down to $H=18.5$ mag
(Andreon \& Pell\'o 2000), which roughly correspond to $R\sim21$ mag or
$M_R\sim-14$ mag. 

At $R\sim16$ mag there is an hint of a possible dip in the LF: $\sim 5$ Coma
galaxies are expected in the half magnitude bin, while $\sim0.7$ is
observed. However, the statistical significance of the effect is negligible
($\sim 1\sigma$). This feature is common among the so far determined Coma
LFs: it has been found in the photographic $V$ (Goodwin \& Peach 1997) and
$b$ band (Biviano et al. 1995), and in the near--infrared  $H$ band (Andreon
\& Pell\'o 2000).

\subsection{Comparison to the literature}

The shaded regions in Figure 3 delimit the best previous determinations of
the LF. In the $R$ band, the shaded region is the LF of the ``deepest and
most detailed survey covering [{\it omissis}] a large area" (Trentham
1998).  He surveyed a $\sim0.18$ deg$^2$ area of the Coma cluster, i.e. a
40 \% smaller area than the present survey, overlapping but not coincident
with the Coma cluster region studied in this paper. At $R>21$ mag,
($M_R>-14.5$ mag) the {\it literature} LF is quite noisy and does not
constrain the LF. The bright part of the Trentham $R$ LF disagrees with
those computed from surveys of large number of clusters (e.g. Paolillo et
al. 2001, Piranomonte et al. 2001), while our LFs are truncated because we
removed giant galaxies and their surrounding area where
the GC blends contamination is potentially high. The
two LFs are normalized to $R\sim18$ and show reasonable agreement, given
the errors, in the region of validity of both LFs. We notice that
errorbars in Trentham (1998) are, in our opinion, underestimated because
they count only once background fluctuations, instead of two as we
advocate. The much shallower $R$ band LF of the Coma cluster computed by
Secker \& Harris (1996) shows a similar agreement. 

In the $V$ band, no LF, comparable in depth and extension to the present
one, is known to the authors.

In the $B$ band, Trentham (1998b) summarizes our present knowledge on the
LF by computing the composite cluster LF, averaging over almost all
literature LFs based on wide field deep images, including Virgo (Sandage,
Binggeli \& Tammann 1985), for example. The shaded region in the bottom
panel shows his result, once data are sampled at 1 mag bins (which help in
reducing the scatter) and vertically shifted to match our points at $B>16$
mag. Our data agree well with the Trentham (1998b) composite LFs and the
agreement should increase if Trentham's (1998b) errorbars were made larger
in order to include twice the background
fluctuations in the error budget.

To summarize, we compute the Coma LF in three bands, over a very large
magnitude range (up to 11 mag) with good statistics. Our results agree
with previous LF determinations on the common magnitude range. The
discussion of the LF is deferred after the presentation of the bi--variate
LF.

\subsection{The bi--variate luminosity function}

The quality of our LF determination allows a truly new interesting quantity
to be accurately determined: the bi--variate LF, i.e. the
LF of galaxies of a given central brightness.  Central
brightness is measured on the images (which are convolved by the
seeing disk, whose FWHM correspond to $\sim1$ Kpc to the Coma cluster
distance) in a 0.25 Kpc aperture.  At the time of the submission of
this paper, this determination was the first so
far accurately computed for any environment, to our best knowledge. The
previous larger effort in this direction is presented in de Jong (1996),
with a study of a sample of 86 field galaxies (while our sample includes $\sim
1000$ cluster member galaxies), whose bi--variate LF is ``more of
qualitative than quantitative interest" (de Jong 1996). After this paper was
submitted, two more bivariate LFs (Cross et al. 2001, Blanton et al. 2001)
were submitted for the publication.

Figure 4 shows a 3D view of the $R$ bi--variate LF. 2D views, at fixed
surface brightnesses, are presented in Figure 5 for both $V$ and $R$ filters.
On the latter figures errorbars can be plotted, and therefore the
quality of the bi--variate LF can be appreciated.
Brightness bins are 1 mag arcsec$^{-2}$ wide, except the brightest one,
which is wider to improve the statistic. Reducing the
amplitude of the first brightness bin decreases slightly the 
statistics but does not change significantly the results.

The way brightness is measured limits the luminosity range accessible
to a galaxy of a given brightness: since the central brightness is
measured on a finite area, objects of a given central brightness have a
minimal flux. The hashed regions in Figure 5 mark the regions that could
not be occupied by our objects because of such a minimal flux.
Furthermore, no object of a given central brightness could be fainter than
a star of the same central brightness. An arrow in the plots marks this
magnitude. In this diagram, compact galaxies fall below the arrow.
Therefore, the observational available range for each bi--variate LF goes
from $-\infty$ to the arrow, including these limits. Therefore, in Figure 4 the
region on the right not occupied by galaxies is empty because the rarity
of such type of galaxies, whereas the empty region on the left is devoid
of galaxies because of the way surface brightness is measured. 

At all brightness bins, galaxies occupy a bounded range in luminosity,
and smaller than the whole available range.  Although a distribution
with a finite width is expected, we can now quantify it. The plotted
values shown in Figure 4 and 5 are tabulated in the joined electronic 
table. 

Galaxies of very large size or very flat surface brightness profile (i.e.
near the left end of each bi--variate LF plot) are uncommon. In fact, 
galaxies  with
$m<\mu-4$, i.e. more than 4 mag brighter than their central brightness,
are almost absent in our sample. Furthermore, the bright end moves
toward fainter magnitudes when the central brightness decreases. 

At the faint end, the LFs seem to flatten or turns down. Furthermore, 
the point at
the arrow magnitude, i.e. objects that are as compact as the seeing disk,
is seldom on the extrapolation of points at brighter magnitudes. While the
possible flattening at magnitude slightly brighter than the arrows
magnitude is uncertain, the points below the arrow are systematically
lower than brighter points. The rarity of galaxies at the magnitude
marked by the arrow, when compared to the expected value based on the
trend at brighter magnitudes, is not due to the fact that compact galaxies
are removed in the star/galaxy classification or implicitly supposed not
to exist, because our derivation of the LF does not follow this path,
unlike previous works. The found rarity of compact galaxies means that
most of the galaxies at the Coma distance are extended sources at our
resolution and explains why our LF, which counts also compact galaxies,
agrees with previous works that instead implicitly assume that compact
galaxies do not exist. 

Lacking a {\it field} bi--variate LF\footnote{During the revision of
this paper two bi--variate {\it field} LFs appear. Nevertheless no
comparison with these works can be performed: the 2dF (Cross et al. 2001) 
SLOAN (Blanton et al. 2001) bi--variate LF  use different definitions of 
brightness
than we have adopted, and, furthermore, 2dF adopts an indirect measure of
the galaxy brightness.} it is difficult to say whether the Coma cluster
is effective in harassing LSB galaxies, as advocated by Moore et al.
(1996; 1999), or the bi--variate LF is the same in the two environments and
it tells us more on galaxy formation and evolution in general.


Figure 6 presents the $R$ band LF (the curve), and a linear fit to the LF
of the galaxies of each central brightness (the lines). The best fit
parameters are listed in Table 2. During the fit process, we manually
discarded outliers points, and we arbitrary adopted a linear (in log
units) fitting function. For some LFs a higher order function would be
preferable but at the price of overfitting most of the other LFs. It is
quite apparent that LSBGs galaxies dominate the LF at faint magnitudes,
while high surface brightness galaxies dominate the bright end.  High
surface brightness galaxies ($\mu_0<20.0$ mag arcsec$^{-2}$) have a
shallow LF ($\alpha\sim-1$), while LSBGs galaxies have a steep and fainter
LF (see also Fig 5).  There is a clear trend for a faintening of the LF
going from high surface brightness galaxies to faint and very faint
central brightnesses, also visible directly on the data in Figure 5. There
is also some evidence for a steepening of the LF, in particular when
galaxies of high surface brightness are considered. The trend is still
there even when not considering at all the LF of galaxies with $\mu_R<19.0$ (or
$\mu_V<20.0$ ) mag arcsec$^{-2}$, which is determined on a wider
brightness range than the other bi--variate LFs, or adopting for this bin a
1 mag arcsec$^{-2}$ wide bin as we did for the other brightness bins.
Surface brightness is correlated to luminosity, since galaxies of lower
central surface brightness have often fainter magnitudes (Figures 4), as
typically found in incomplete volume samples (e.g. van der Hulst, et al.
1993; de Blok, van der Hulst \& Bothun 1995; Impey \& Bothun 1997, van den
Hoek et al. 2000).


\section{Discussion}

\subsection{What changes going deeper}

First of all, we emphasize that adopting a total magnitude
instead of an isophotal magnitude should probably modify slightly the LF
and the bi--variate LF by shifting them to the left and making them
steeper. The LF becomes slightly brighter, i.e. moves to the left, because
a fraction of the total flux is below the brightness threshold, thus the
adopted isophotal magnitudes underestimate the total flux of the galaxies.
For galaxies of high central brightness (i.e. bright, see section 4.3) the
fraction of lost flux is fairly small (Trentham 1997) if their surface
brightness profile below the observed brightness threshold follows the
extrapolation of the observed part: the galaxy flux has been
already integrated over a wide brightness range and the flux below the
threshold is negligible. On the other end, the fraction of the total flux
below the brightness threshold increases going toward faint magnitudes
objects because these objects often have faint central brightnesses (sect. 4.3)
and thus their isophotal magnitude is integrated on smaller and smaller
brightness ranges. Assuming that galaxies have perfect exponential surface
brightness profiles, we find that the corrections range from $-0.75$ mag
for the galaxies of lowest surface brightness to $-0.05$ mag for high
surface brightness galaxies. Because of this correction, the $R$ band LF
changes its slope $\alpha$ by $-0.05$, i.e. by 3 \%. The LF likely becomes
steeper when adopting total magnitudes for yet another reason: galaxies
whose central brightness is below the present brightness threshold will be
counted and they are likely preferably faint, if the trend presented in
Figure 5 continues at lower surface brightnesses and magnitudes.

\subsection{Large numbers of LSBGs in high density regions}

We found a large quantity of LSBGs in the core of the Coma cluster.
The cluster environment is often regarded as hostile to the formation and
survival of LSBGs, in fact as much as 90 \% of the stars in LSBGs
can be harassed from them (Moore et al. 1999). On the other end,
the harassment process may contribute to the production of LSBGs in
clusters (Moore, 1996). Therefore, the 'harassment' paradigm has no
predictive power on the number of LSBGs in clusters. Maybe the
cluster LF might tell us about cluster--related processes in a too detailed
level for a prediction with the present day models. 

Phillipps et al. (1998a) examine the dissimilarity of the dwarf population
in different environments. Their faintest dwarfs are 5 mag brighter than
our limit, i.e. they are talking about normal dwarfs, not faint ones. They
note a variation in the LF shape which is driven in part by galaxy
density: at low galaxy densities both steep and shallow LFs are permitted,
while at high galaxy density only flat LFs are observed. We computed,
according to their recipes, the giant--to--dwarf ratio (which in our case
used only galaxies with S/N$>\sim300$) and found a giant to dwarf ratio of
$\sim7\pm1$, at the projected galaxy density of 26 gal Mpc$^{-2}$. The
result is near the extrapolation of the outer envelope of their proposed
correlation. This calculation is computed using $H_0=50$ km s$^{-1}$ Mpc
for consistency with Phillips et al. (1998a).

\subsection{Missed galaxies in the field LF determination?}

There is ample discussion in the literature whether or not the local field
LF is well determined at faint magnitudes, because most of the surveys
purport to be magnitude limited do not take into account surface
brightness effects (Disney 1997; Sprayberry et al. 1997; Phillipps et al.
1998b). Optical surveys reveal an excess of faint blue galaxies over and
above the number predicted by simple models relating local to distant
observations (e.g. Tyson 1988; Lilly et al. 1991). The flat faint--end
slope measured in the local $B$--band LF of galaxies (Efstathiou et al.
1988; Loveday et al. 1992) plays an important role in this interpretation,
for the faint blues galaxies might otherwise be explained by a local
population of intrinsically faint (and nearby) galaxies (Driver \&
Phillipps 1996). 

Recent surveys, such as the cluster survey by Impey, Bothun \& Malin (1988)
and Irwin et al. (1990), and the field survey by Impey et al. (1996) have
taken into account this potential source of bias by deliberately searching
for LSBGs. However, their search is limited to giant LSBGs,
i.e. dwarfs LSBGs are not sampled at all. Some of these works also use 
galaxies whose size or surface brightness is near the survey limits and are
obliged to statistically correct their sample for missed galaxies by
adopting simplifying assumptions, for example that LSBGs have perfect
exponential surface brightness profiles. The same assumption is done again
in computing the volume correction (for field surveys). 

Our own sample is a bit different from previous surveys: first of all, it
is a volume--limited sample, since it is a cluster sample.  Unlike
previous cluster surveys, we do not impose a large minimal size (say 30
arcsec as Impey, Bothun \& Malin 1988 for Virgo candidate galaxies), but
we select the sample by absolute magnitude. Therefore, {\it dwarf}
LSBGs are not discarded {\it ab initio} by adopting a large angular
diameter for galaxies, provided that their flux brighter than the
isophotal threshold is larger than the magnitude of completeness.
Furthermore, previous LSBGs searches compute the LSBG contribution to the
LF {\it assuming} that all detected LSBGs belong to the studied cluster,
while we compute the background and foreground contribution by using a
control field. With respect to field LSBGs searches, the advantages of the
present determination are even larger: first of all, our sample is,
as explained, a magnitude complete sample, while field sample are often
diameter selected, at a large angular diameter (for example, the survey of
O'Neil, Bothun \& Cornell (1996) that have a similar depth uses
a 143 arcsec$^2$ minimal size, but at a 2-3 mag deeper isophote).
Second, we choose to work only with the high S/N part of the catalog, thus
completely skipping the problem of the correction for missed detections
near the survey limits (minimal size and minimal brightness). Most
importantly, the sample is volume--complete, and no volume
correction/selection function should be computed since the visibility of
Coma galaxies does not depend on the redshift.
It is true that beside Coma LSBGs there are other LSBGs in the Coma
line of sight, but these are removed statistically from the sample. We
remind the reader that the calculation of the volume correction/selection
function for a diameter+brightness selected (field) sample is so difficult
that many experienced astronomers, including Disney (1976), got it
wrong when computing it (Disney 1999). This correction is quite large and
thus uncertain. For example, the median incompleteness correction applied
in the calculation of the LF of LSBGs by Sprayberry et al. (1997), is
five, which means that the one detected object has been used to infer the
presence of four other galaxies escaping detection or redshift
determination with similar photometric parameters and in the same universe
volume. 



Sprayberry et al. (1997) find a steep LF for LSBGs, steeper than the LF usually
found in samples claimed to be flux--limited. Their LF is computed in the $B$
band and concerns galaxies with $\mu_B(0)>22$ mag arcsec$^{-2}$.  Assuming that
field LSBGs have an average $B-V=0.5$ mag (e.g. de Blok et al. 1995) we can
compute the $V$ band field LF for galaxies with $\mu_V(0)>21.5$ mag
arcsec$^{-2}$ from the $B$ LF. The latter cut in surface brightness is applied
to the Coma galaxies in order to compare the two LFs. Figure 7 compares the
result of this exercise. There is a remarkable agreement on the location and
slope of the exponentially decrease of the two LFs. Instead, 
the two LF are arbitrary shifted. The minor difference at $V<17.5$ mag concerns
3.7 galaxies missed in the present LF, that can be fully accounted for by
statistical fluctuations and by the difference in the passband used for
selection. If the LSBGs LF is independent of the environment, this agreement
confirms the correctness of the Sprayberry et al. (1997) calculation of the
visibility function (actually his survey has a selection function even more
complex than those of diameter+brightness limited surveys) and gives support to
their claim that local field surveys overlook a numerous population of LSBGs.

%
%

\section{Conclusion}

Wide field images of a nearby cluster, coupled with a exhaustive analysis of
the sources of error, allow us to extend the LF to very faint magnitudes and
to include the LSBGs contribution. Most importantly, the data allow the
determination of the bi--variate LF, without missing
LSBGs (down to a faint limiting central brightness) or losing compact galaxies,
because of their 
resemblance to stars or to background galaxies. The present bi--variate
LF determination has a straightforward selection function allowing a precise
measure of the frequency defining how galaxies occupy the available space in
the central surface brightness $vs$ magnitude plane in the Coma cluster.
Furthermore, the present determination does not need uncertain corrections
for passing from the observed distribution to the actual galaxy
distribution, simply because the sample is naturally volume--limited, or
uncertain assumption on the membership of faint galaxies, because
foreground and background has been statistically removed. LSBGs are 
by far the largest galaxy population, most of them are also quite faint and
this study suggests that we have not yet reach the magnitude or the central
brightness turn off (if it exists). On the other hand, compact galaxies are a
minority population.

\acknowledgements  This work has been completed thanks to the efforts of
many peoples involved in several projects: the CFH12K team provided a camera
that worked smoothly right from the night of its first light, Bill Joye from
SAO provided ds9, an efficient viewer for these large field images, 
Emmanuel Bertin provided SExtractor, an efficient detection and
classification software optimized for large images.  The authors wish to
thank the CFHT director P. Couturier for the allocation of CFH12K
discretionary time, M. Cr\'ez\'e and A. Robin for providing their 1998 raw
UH8K data of the SA\,57 field and G. Fahlman for his attentive lecture of 
this paper. S.A. thanks Prof. Guido Chincarini, for their
kind hospitality, at Osservatorio Astronomico di Brera, were part of this
paper has been written, and Prof. Massimo Capaccioli, director of the
Osservatorio Astronomico di Capodimonte for allowing a long stay there.

\newpage
 
\begin{deluxetable}{lccccccl}
\tablecolumns{8}
\tablewidth{0pc}
\small
\tablecaption{The sample}
\tablehead{
\colhead{Pointing} & \colhead{filter} & \colhead{$t_{exp}$} &
\colhead{Instrument} & \colhead{seeing (FWHM)\tablenotemark{a}} &
\colhead{detec. $\mu$\tablenotemark{a}} &
\colhead{useful area}\tablenotemark{b}\\
& & \colhead{sec} & & \colhead{arcsec} & mag arcsec$^{-2}$ & degree sq.
}

\startdata
Coma & B & 180 $\times$ 4 & CFH12K & 0.88 & 25.0 & 0.20 \nl
Coma & V & 180 $\times$ 4 & CFH12K & 1.23 & 25.5 & 0.29 \nl
Coma & R & 120 $\times$ 4 & CFH12K & 1.04 & 24.5 & 0.29 \nl
NGC 3486 & B & 600 $\times$ 14 & CFH12K & 0.72 & $ 26.5 $ & 0.17 \nl
SA 57 & V & 1200 $\times$ 8 & UH8K & 0.65 & $ 26.2 $ & 0.18 \nl
SA 57 & R & 1200 $\times$ 8 & UH8K & 0.65 & $ 25.7$ & 0.18 \nl
\enddata
\tablenotetext{a}{Before matching observations to Coma data}
\tablenotetext{b}{After removing areas noisier than average, and halos of bright galaxies}
\end{deluxetable}

\begin{deluxetable}{lcllllll}
\tablecolumns{8}
\tablewidth{0pc}
\small
\tablecaption{Best fits}
\tablehead{
\colhead{} & \colhead{range} & \colhead{0th ord. coeff.} & \colhead{1st ord. coeff.} & 
\colhead{2nd ord. coeff.} & \colhead{3rd ord. coeff.} & \colhead{$\chi_{\nu}^2$}}
\startdata
LF--R & 13.00--23.25 & 28.66 & -5.1346 & 0.30346 & -0.005672398 & 1.3 \nl
LF--V & 13.25--23.75 & 26.73 & -4.7455 & 0.27751 & -0.005130364 & 0.7 \nl
LF--B & 14.75--22.25 & 27.76 & -4.4728 & 0.24188 & -0.004187621 & 1.0 \nl
\hline
LF--R, $\mu<19$ & 12.75--17.75 & 1.80 & -0.08 & \nodata & \nodata & \nodata \nl
LF--R, $\mu=19.5$ & 15.5--19.0 & -2.13 & 0.17 & \nodata & \nodata & \nodata \nl	
LF--R, $\mu=20.5$ & 16.5--20.0 & -4.72 & 0.33 & \nodata & \nodata & \nodata \nl
LF--R, $\mu=21.5$ & 17.5--21.5 & -5.92 & 0.38 & \nodata & \nodata & \nodata \nl
LF--R, $\mu=22.5$ & 18.5--22.25 & -7.21 & 0.43 & \nodata & \nodata & \nodata \nl
LF--R, $\mu=23.5$ & 19.5--23.0 & -18.83 & 0.48 & \nodata & \nodata & \nodata \nl 
\enddata
\tablecomments{The quoted coefficients are a simple empirical description of the
LF shape and the large number of digits should not be taken as an indication
of the good quality of the fit. Furthermore, the fit should not extrapolated
outside the quoted range of validity.}
\end{deluxetable}
 
\newpage
\begin{figure*}
\centerline{\psfig{figure=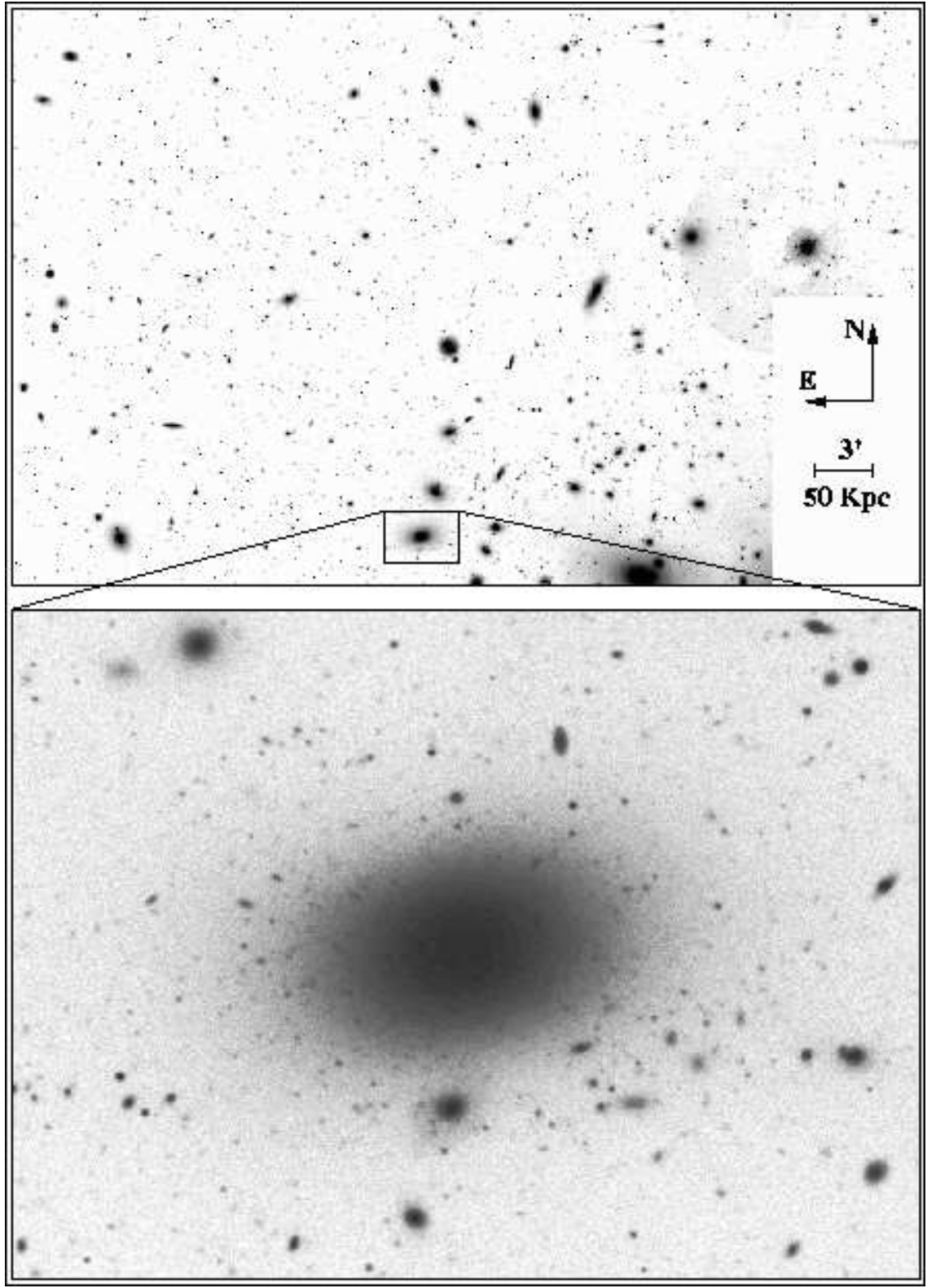,width=15truecm,angle=0}}
\caption[h]{
The top image shows the whole CFH12K field of view 
$R$ image of the studied field. North is up and east is to the left. 
The field of view is $42\times28$ arcmin$^2$, i.e. $1.2\times0.8$ Mpc$^2$
at the Coma distance. Regions with lower quality than average are not considered
(such as the bottom right CCD).
The studied $B$ field includes the central square area.
The bottom image shows the galaxy IC\,4051, dwarfs and several
GCs blends. A compressed figure is included in ASTRO-PH.}
\end{figure*}

\newpage
\begin{figure}
\epsfxsize=10cm
\centerline{\psfig{figure=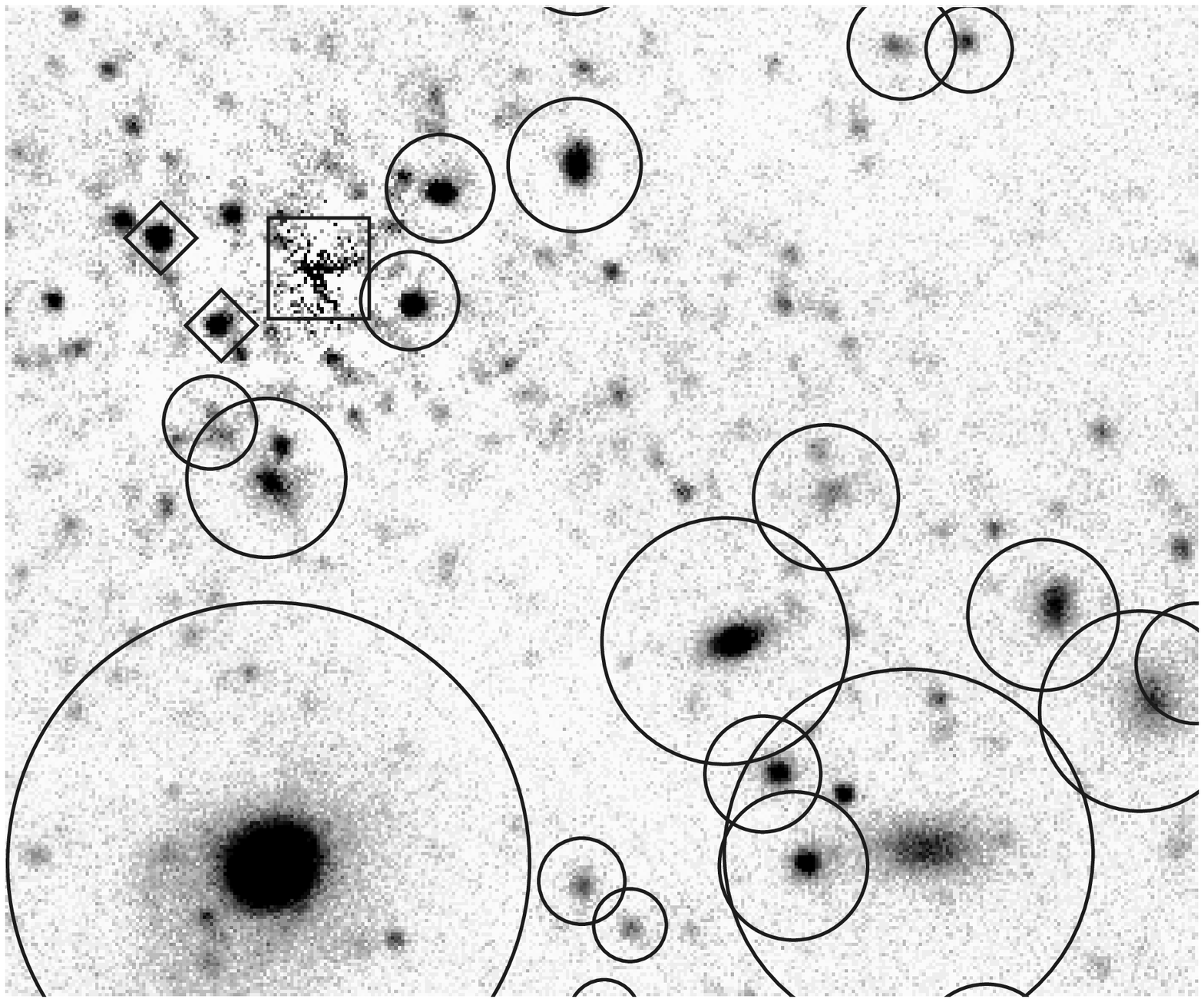,width=7truecm}%
\psfig{figure=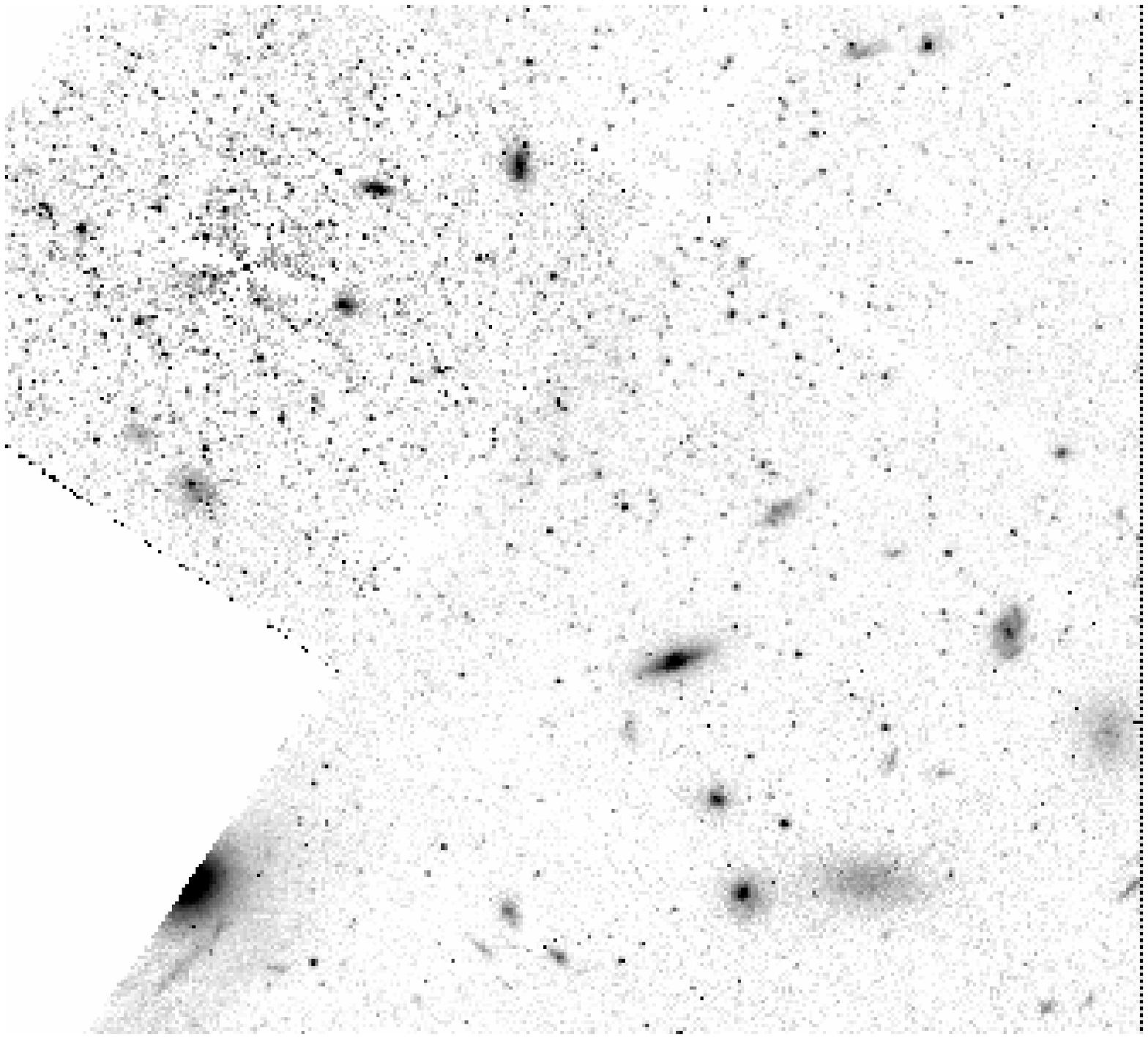,width=7truecm}}
\caption[h]{Residual $R$ image after having subtracted a model of
IC\,4051 from the original image. The left panel shows a part of the ground
image, while the right one displays a part of the {\it HST} one. 
The galaxy, shown in the low panel of Figure 1, is much larger than the
field of view of this cutout of 68 arcsec angular size. In the left
panel, the square marks IC\,4051's center, and circles the true galaxies
in the field, as confirmed by the {\it HST} image. Most of the remaining 
objects, most of which looks as extended sources from the ground
and are brighter than the  completeness magnitude, are unresolved 
blends of GCs.  Diamond points mark the two brightest blends in the {\it HST} 
field of view, having $R\sim21$ mag.}
\end{figure}

\newpage

\begin{figure*}
\hbox{
\epsfysize=6.5cm
\epsfbox[45 195 560 540]{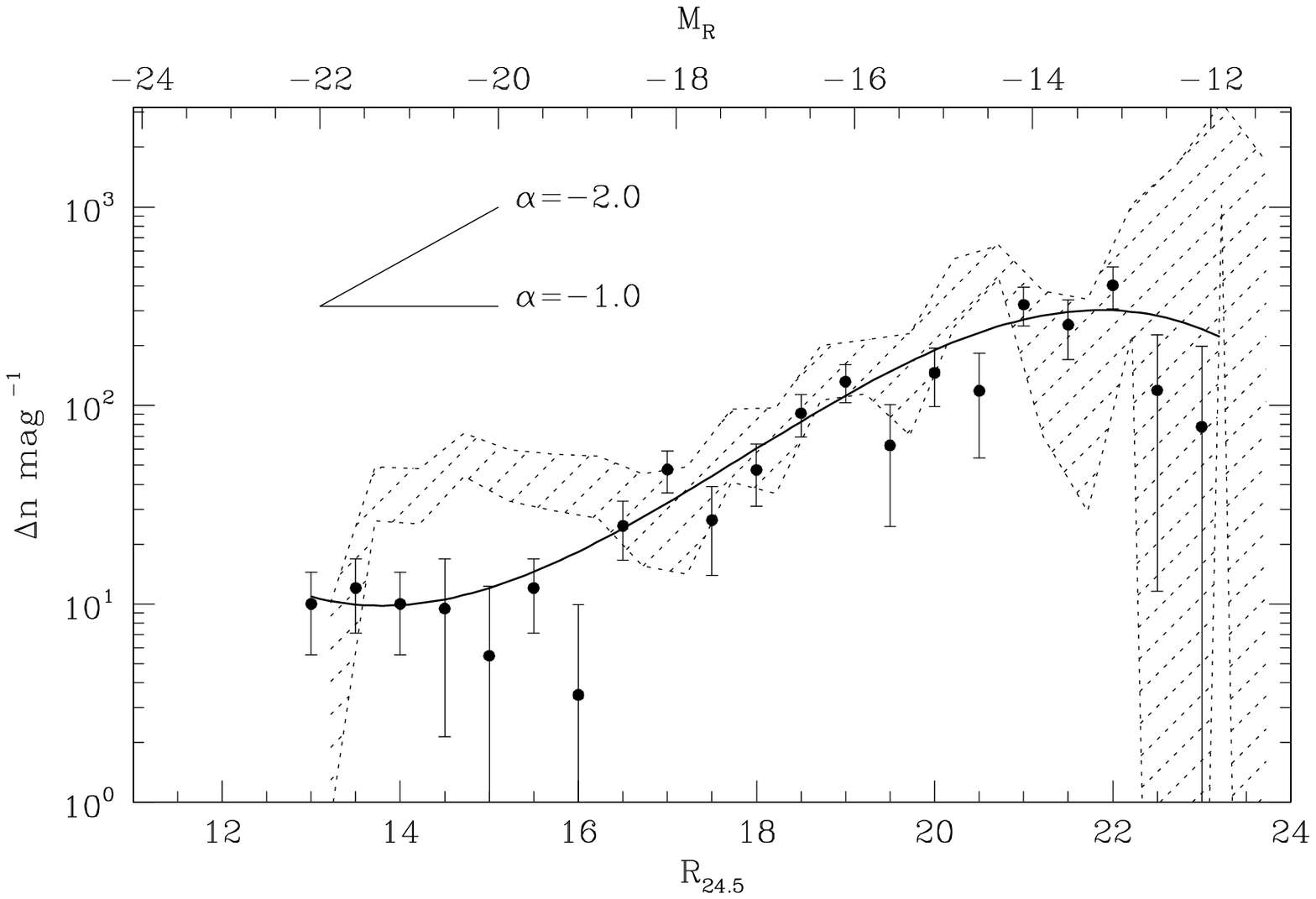}
\hfill\null}
\hbox{
\epsfysize=6.5cm
\epsfbox[45 195 560 540]{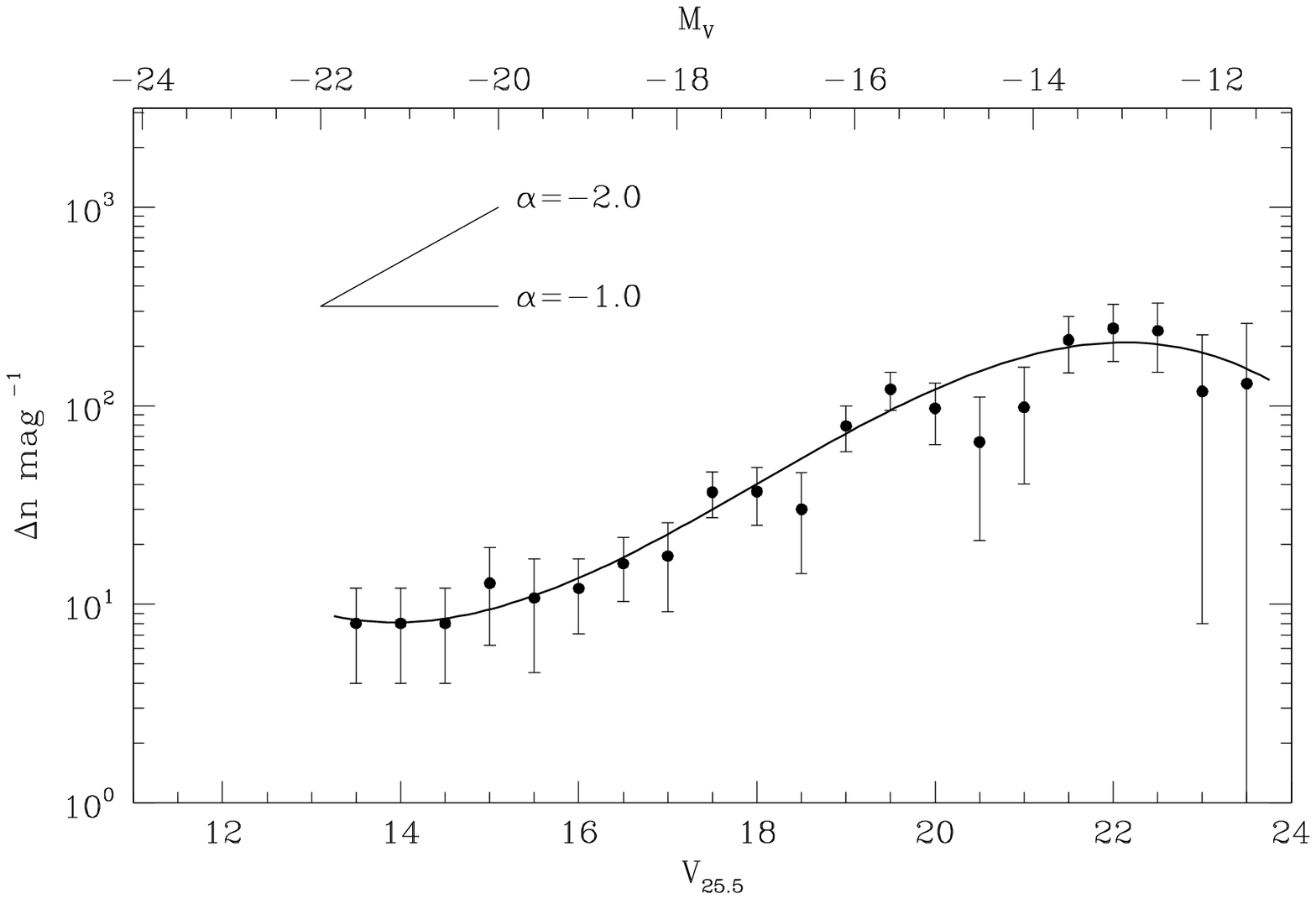}
\hfill\null}
\hbox{
\epsfysize=6.5cm
\epsfbox[45 195 560 540]{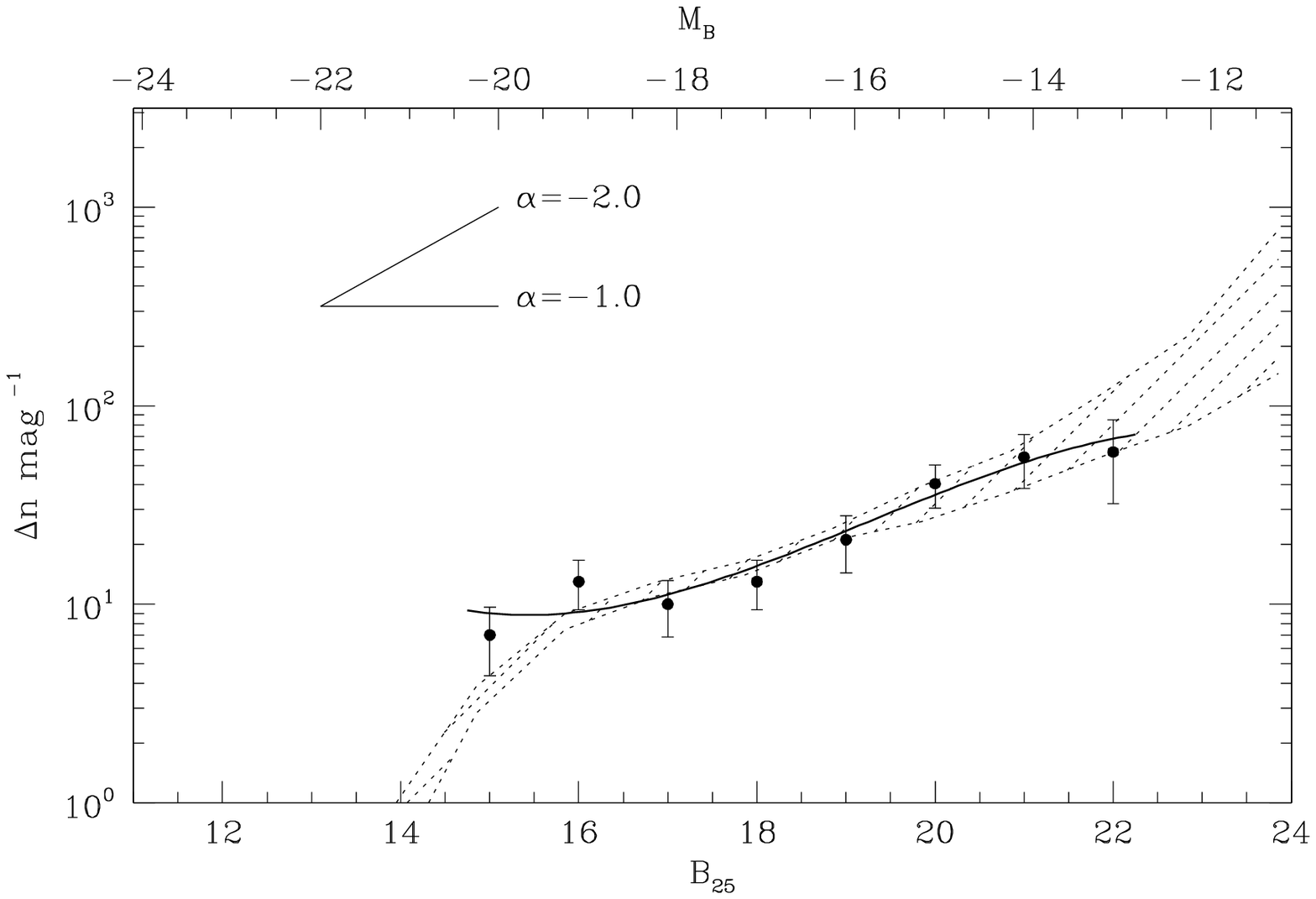}
\hfill\null}
\caption[h]{Luminosity function of Coma cluster in the $R$, $V$ and $B$ bands
(close points). Both apparent and absolute magnitudes are shown on the
abscissa. Errorbars take in full account Poissonian and non--Poissonian
errors, i.e. include errors due to the presence of under/overdensities
along the lines of sight. The thick curve is the best fit with a 3 degree 
power--law to the data. In the $R$ (top) panel, the hashed region delimit
the LF determined by Trentham (1998a). 
In the $B$ (bottom) panel, the hashed region
delimit the composite $B$ LF, averaged over almost all literature
ones (from Trentham 1998b).}
\end{figure*}

\newpage
\begin{figure*}
\centerline{\psfig{figure=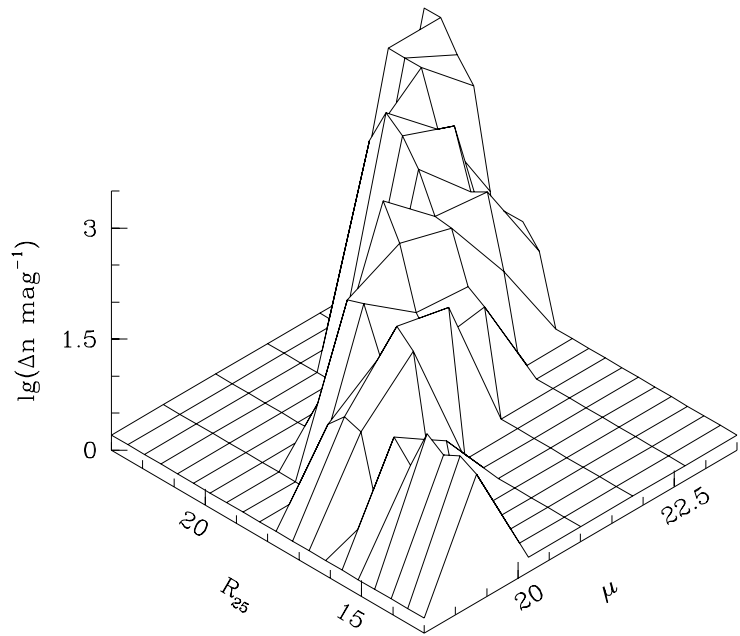,width=15truecm}}
\caption[h]{3D view of the bi--variate LF of Coma galaxies in the $R$ band.
The empty region on the left is devoid of galaxies because of the way brightness
is defined, while the region on the right is empty because the rarity of such
a type of galaxies in Coma.}
\end{figure*}

\newpage
\begin{figure*}
\centerline{\psfig{figure=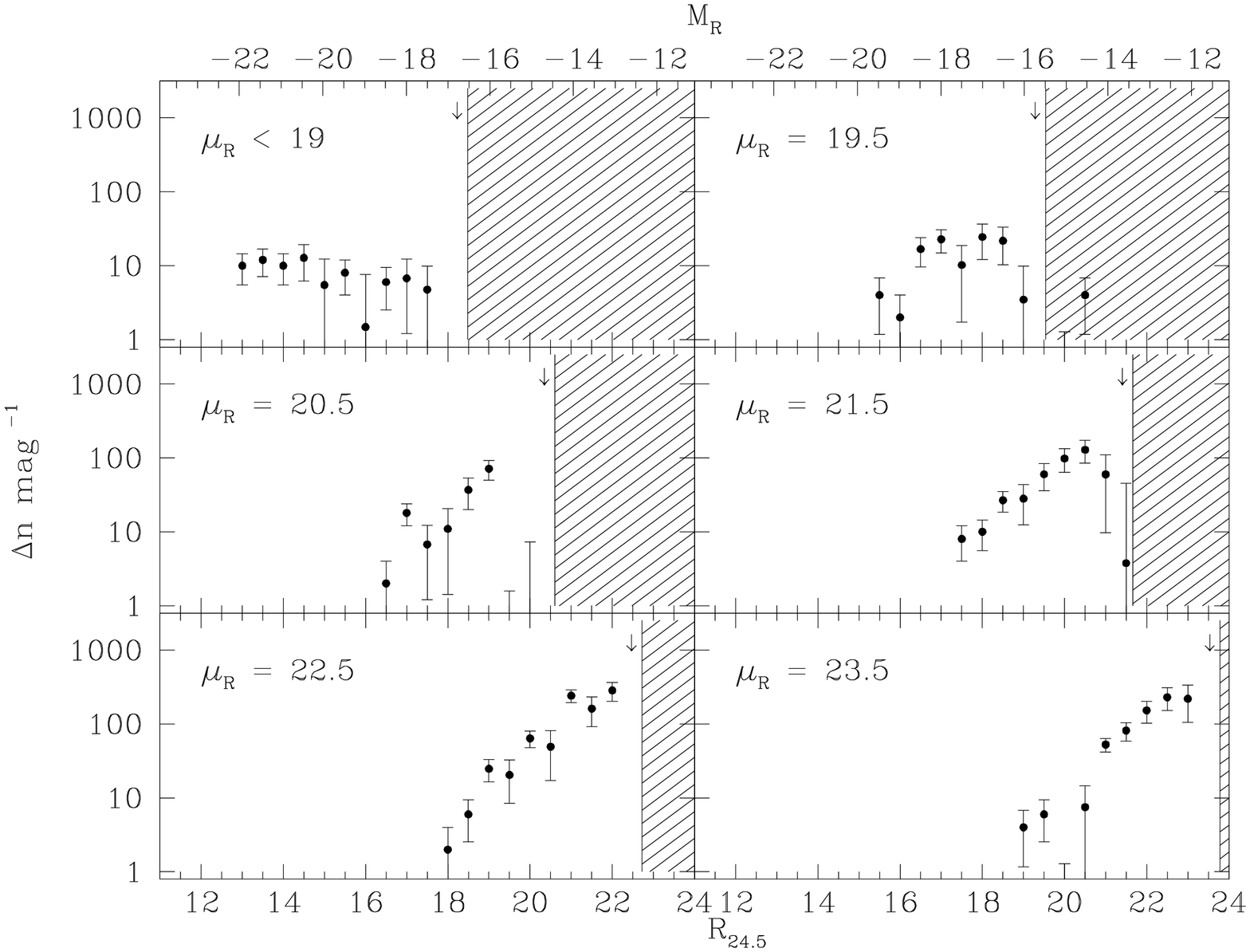,width=15truecm,angle=-90}}
\centerline{\psfig{figure=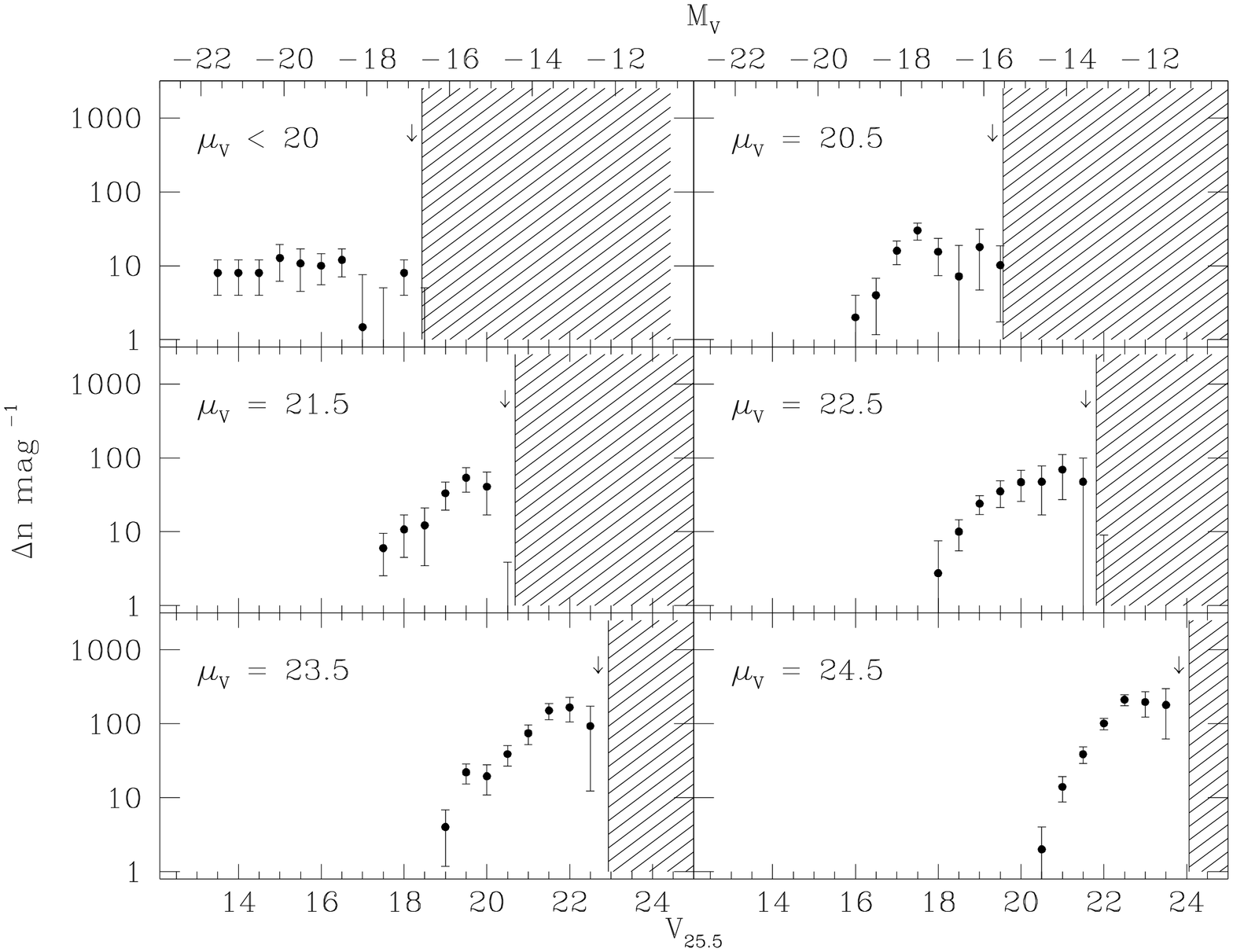,width=15truecm,angle=-90}}
\caption[h]{Bi--variate LF of Coma galaxies in the $R$ and $V$ bands. 
There is a clear progression from flat and bright LFs of HSB galaxies 
to steep and faint LFs of LSB. Errorbars are as in Figure 3.
Seeing and sampled area for the brightness determination make
the hashed part of the diagram forbidden to galaxies.}
\end{figure*}

\begin{figure*}
\hbox{
\epsfysize=8cm
\epsfbox[45 195 560 540]{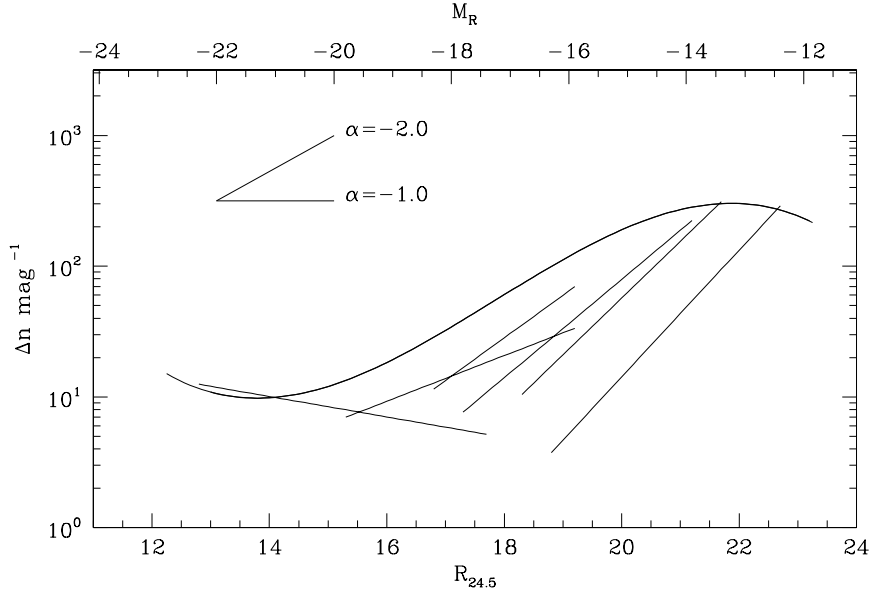}
\hfill\null}
\caption[h]{Dissection of the $R$ LF in central brightness. The smooth
curve is the Coma $R$ band LF, while the straight lines are the contribution to
the LF of the galaxies of each central brightness. Galaxies of large
central brightness have flat LF (on the left part of the graph),
while LSB have steep LF (on the right). Bins of central brightness are:
$<19, 19.5, 20.5, 21.5, 22.5, 23.5$ mag arcsec$^{-2}$.
}
\end{figure*}

\begin{figure*}
\hbox{
\epsfysize=8cm
\epsfbox[55 195 380 540]{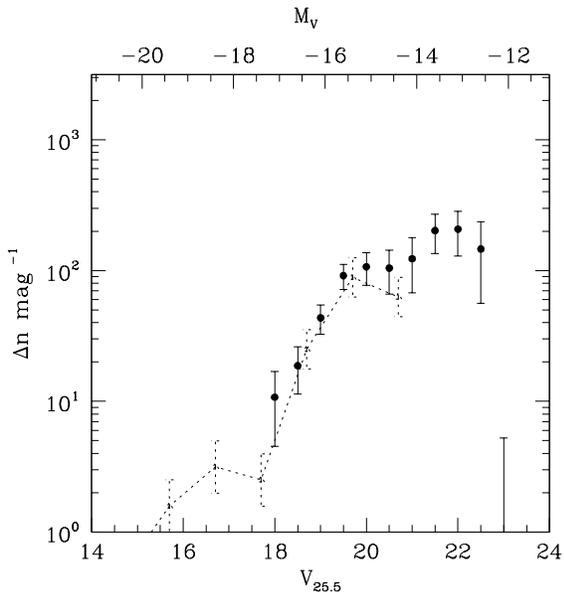}
\hfill\null}
\caption[h]{Comparison LF of LSBGs field galaxies
(dotted spline and errorbar, taken from Sprayberry et al. 1997), 
transformed from the $B$ band (see text for details), to the 
LF of Coma LSBGs galaxies (filled points and solid errorbars). 
The amplitude of the field LF has been vertically shifted to match the
much denser Coma cluster.
}
\end{figure*}
\end{document}